\documentclass[11pt]{article}

\usepackage[margin=1in]{geometry}
\usepackage[T1]{fontenc}
\usepackage{times}
\usepackage{booktabs}
\usepackage{longtable}
\usepackage{array}
\usepackage{enumitem}
\usepackage{graphicx}
\usepackage{multirow}
\usepackage{hyperref}
\usepackage{amsmath}
\usepackage{amssymb}
\usepackage{xcolor}
\usepackage{url}
\usepackage{tikz}
\usetikzlibrary{arrows.meta,positioning,calc,shapes.geometric}

\hypersetup{
colorlinks=true,
linkcolor=black,
citecolor=black,
urlcolor=blue
}

\title{
Rotary GPU:\\
Exploring Local Execution Paths for Large Mixture-of-Experts Models Under Limited GPU Memory
}

\author{
Myeong Jun Jo\\
ANIMA Research, Independent Researcher, South Korea\\
ORCID: \href{https://orcid.org/0009-0006-9540-4666}{0009-0006-9540-4666}
}

\date{June 2026}

\begin{document}

\maketitle

\begin{abstract}
Large language models have achieved remarkable capabilities through scaling, and this paper does not challenge that. It instead investigates a different question: once large models already exist, can they become more accessible to environments with substantially smaller hardware resources? The motivation came from deployment concerns rather than architecture research. Many organizations operate under hardware, budget, security, or closed-network constraints that limit access to large accelerator clusters, and as models continue to improve, deployment accessibility may matter as much as capability itself.

This paper presents Rotary GPU, an exploratory execution approach derived from a previously disclosed rotary-based accelerator residency concept. A public validation was conducted using a Qwen3.6-35B-A3B-class Mixture-of-Experts model executed locally on a consumer laptop with an RTX 4060 Laptop GPU containing 8 GB of VRAM. Under the primary configuration, the system generated 2048 output tokens while maintaining approximately 6.3 GB of VRAM usage and an observed decode throughput of 21.06 tokens per second. The goal is not to replace data-center infrastructure but to explore whether some capabilities of large models can be brought closer to environments where such infrastructure is unavailable. The results should be read as exploratory rather than definitive, but they suggest deployment accessibility deserves continued investigation as these models evolve.
\end{abstract}

\section{Introduction}

Since the Transformer architecture and the subsequent success of the GPT family, progress in language modeling has been strongly associated with scale. Larger datasets, larger models, and more powerful infrastructure have repeatedly produced stronger performance, and modern systems now summarize documents, generate software, translate, and answer questions at a level that would have seemed unrealistic only a few years ago. The success of this scaling paradigm is difficult to dispute, and the purpose of this work is not to argue against it.

Instead, it asks a different question: once large models already exist, must they always remain coupled to equally large hardware? Most discussion of deployment treats additional memory, larger accelerators, and larger serving clusters as the natural consequence of more capable models. That approach has been remarkably successful. But deployment reality is not identical to research reality. Many organizations operate under practical constraints involving security, compliance, hardware availability, infrastructure cost, physical space, and operational complexity. Government agencies, financial institutions, manufacturing facilities, healthcare organizations, and defense-related environments frequently maintain closed or partially isolated networks where access to large-scale cloud infrastructure may be restricted even when advanced model capabilities are genuinely wanted.

This observation motivated the central question of the paper. If only a subset of a model appears relevant to a particular context at a given moment, must the entire model always remain resident in accelerator memory? A simple analogy captures the intuition: a warehouse remains valuable even when a customer requests only a single item, and the real question is not whether the warehouse should exist but whether every delivery requires moving the whole warehouse along with the package. Most delivery systems keep the warehouse where it is and move only what is likely to be needed. Large-model inference is far more complex than package delivery and the analogy should not be taken literally, but it is the intuition that started this investigation.

The same idea appears across many large information systems. At any given moment only a subset of available information is actively relevant, and retrieval systems, memory systems, and routing systems all exploit this. This work explores whether accelerator residency management might benefit from a similar perspective. The idea grew out of a broader line of independent investigations by the author into rotary-based structures for retrieval, memory organization, routing, and execution management, extended here into the domain of local large-model execution. Rather than disclosing implementation details, the paper focuses on public validation and externally observable behavior.

The objective was not to establish state-of-the-art performance or to replace existing deployment architectures. It was considerably simpler: to investigate whether meaningful execution of a large Mixture-of-Experts model could remain possible under hardware conditions traditionally regarded as insufficient for models of that scale. The results should therefore be interpreted as exploratory. Additional hardware platforms, operating systems, workloads, and independent reproductions remain necessary before broader conclusions can be drawn, but the observations suggest that deployment accessibility deserves investigation as a problem partially independent from model scaling itself.

\section{Motivation and Deployment Reality}

The motivation behind Rotary GPU emerged from deployment concerns rather than from an attempt to redesign language-model architecture. My professional background is in organizational operations and human resource management, where much of the work involves information access, repetitive tasks, process efficiency, document handling, and operational decision-making. As language models became more capable, a natural question emerged: how can these capabilities become accessible inside ordinary organizations?

The answer is not obvious. Many organizations do not possess dedicated AI infrastructure, and even when advanced models could provide clear value, deployment faces practical constraints involving hardware budgets, procurement processes, security requirements, maintenance responsibilities, and network restrictions. A large cloud-hosted model may offer exceptional performance yet remain impractical if organizational policy prevents sensitive information from leaving internal infrastructure. At the same time, building large accelerator clusters is rarely feasible for such organizations, because hardware acquisition, electrical requirements, cooling, maintenance cost, and operational expertise all represent real barriers. The result is that many organizations will deploy smaller systems even when larger models would be preferable.

This reframes the problem. Instead of asking how to continuously increase available hardware, one can ask whether deployment strategies themselves can become more efficient: can larger models become usable within smaller hardware envelopes, and can local execution become practical for organizations that cannot maintain large AI infrastructure? Rotary GPU emerged from these questions. The objective is not to replace cloud systems, which remain indispensable for training frontier models and supporting large-scale services. Accessibility, rather than replacement, is the motivation.

\section{Related Work}

The foundation of modern language modeling was established by the Transformer architecture, which introduced attention-based sequence modeling and enabled substantial improvements in scalability. Subsequent GPT-family systems demonstrated that increasing model size, dataset size, and training computation could produce consistent improvements across a broad range of tasks. As scale increased, inference efficiency became an increasingly important research area, and numerous approaches now exist to reduce memory consumption or improve throughput, including quantization, activation sparsity, memory compression, speculative decoding, KV-cache optimization, accelerator-aware scheduling, and parameter offloading. Quantization in particular has become one of the most practical techniques for local deployment and is widely used on consumer hardware.

Mixture-of-Experts architectures introduced another important observation. Rather than activating every parameter for every token, MoE systems route computation through a subset of available experts, allowing access to larger parameter spaces without proportional increases in per-token cost. This selective computation naturally raises a question about selective residency: if only a subset of experts participates in computation at a given moment, must every expert remain continuously resident in accelerator memory? Several existing systems investigate related ideas through memory offloading, hierarchical storage, prefetching, host-device coordination, and hybrid execution, demonstrating that residency can be treated as a dynamic resource-management problem rather than a static allocation problem.

The present work explores a different angle. Instead of viewing residency solely as cache management, Rotary GPU investigates whether residency decisions might evolve through structured rotary transitions influenced by execution context. The purpose is not to claim superiority over existing approaches but to present an exploratory validation showing that large-model execution may remain possible under substantially constrained hardware.

\section{Rotary GPU Concept}

Rotary GPU rests on a simple observation: during inference, not every component of a large model contributes equally at every moment. Certain experts, routing paths, execution regions, or parameter groups become highly relevant within a particular context while others remain inactive. The complete model remains important overall, but immediate computational demand tends to concentrate within a smaller subset of resources. This motivates a residency-management perspective in which, instead of assuming every component must remain permanently resident, the system prioritizes components likely to become relevant in the near future.

Conceptually, Rotary GPU treats accelerator residency as a rotating resource-management problem. Residency locations are treated as slots rather than fixed assignments, and rather than assuming that every component remains continuously resident, residency candidates are allowed to move between slots according to structured rotational scheduling decisions. The concept can be understood through three principles: residency is dynamic rather than permanent; residency decisions may be influenced by information generated during execution rather than solely by static configuration; and residency transitions may follow structured cyclic patterns rather than purely reactive replacement policies. This stands in contrast to conventional Least-Recently-Used eviction, which advances in one direction by usage recency alone and provides no structured way to return to a previously resident set when an earlier context recurs.

Inventory management offers an intuitive parallel. A warehouse containing thousands of items does not require every item at the loading dock simultaneously; inventory moves between storage locations according to expected demand. The point of Rotary GPU is not to decide whether parts of the model are important, because the complete model remains important. The question is whether every component must remain equally accessible at all times.

Figure~\ref{fig:arch} shows the overall structure and Figure~\ref{fig:rotation} illustrates the cyclical update of the slot group. This paper does not disclose implementation details, scheduling algorithms, protection mechanisms, or internal execution logic; its purpose is to evaluate whether the broader architectural direction can produce meaningful execution outcomes under constrained hardware, which would matter for deployment scenarios where available accelerator memory is substantially smaller than the apparent size of the target model.

\begin{figure}[t]
\centering
\begin{tikzpicture}[
  font=\footnotesize,
  box/.style={draw,rounded corners=1pt,align=center,inner sep=4pt},
  hostbox/.style={box,fill=black!4},
  slot/.style={draw,circle,minimum size=0.85cm,inner sep=0pt,font=\scriptsize},
  >={Stealth[length=2mm]}
]
\node[hostbox,minimum width=4.4cm] (hmem) at (0,0) {\textbf{Host Memory / Storage Tier}\\\scriptsize full model weights (footprint $>$ VRAM)};
\node[box,minimum width=4.4cm,below=0.7cm of hmem] (sub) {\textbf{Sub-modules} (experts / layers)\\[2pt]
  \scriptsize\begin{tabular}{cccc} SM$_1$ & SM$_2$ & SM$_3$ & $\cdots$\\ SM$_{n-1}$ & SM$_n$ & & \end{tabular}};
\node[box,minimum width=4.4cm,below=1.5cm of sub] (route) {\textbf{Routing Result}\\\scriptsize hidden state $h$, routing vector $r$};
\node[hostbox,minimum width=5cm,right=2.4cm of hmem,yshift=-0.1cm] (slots) {\textbf{Rotary Slot Group}\\[10pt]
  \tikz[baseline]{
    \node[slot] (s1) at (0,0) {slot 1};
    \node[slot] (s2) at (1.2,0) {slot 2};
    \node at (2.4,0) {$\cdots$};
    \node[slot] (s3) at (3.6,0) {slot N};
    \draw[->,bend right=35] (s1) to node[below,font=\scriptsize\itshape]{cyclical rotation} (s3);
  }};
\node[box,minimum width=5cm,below=0.7cm of slots] (lut) {\textbf{Lookup Table}\\\scriptsize sub-module ID $\to$ slot position};
\node[box,minimum width=5cm,below=0.5cm of lut] (ctrl) {\textbf{Rotary Controller}\\\scriptsize rotation transform $\to$ rotary projection};
\draw[->,dashed] (hmem.east) -- node[above,font=\scriptsize]{selective load} (slots.west);
\draw[->,dashed] (slots.west|-sub.north) -- node[below,font=\scriptsize,yshift=-1pt]{evict} (sub.east);
\draw[->] (route.east) -- (ctrl.west);
\draw[->] (ctrl.north) -- (lut.south);
\node[font=\scriptsize\itshape,text width=11cm,align=center,below=0.5cm of route,xshift=2.7cm]
  {Only a subset of sub-modules resides on the accelerator at any moment; residency is updated by rotation rather than by simple LRU eviction.};
\end{tikzpicture}
\caption{Rotary GPU residency system. Full model weights remain in host memory while only a rotating subset of sub-modules resides on the accelerator. Redrawn from KR Patent Publication 10-2026-0070380, Fig.~1.}
\label{fig:arch}
\end{figure}

\begin{figure}[t]
\centering
\begin{tikzpicture}[
  font=\footnotesize,
  slot/.style={draw,circle,minimum size=1.0cm,inner sep=0pt,font=\scriptsize},
  >={Stealth[length=2.4mm]}
]
\def\R{2.6}
\foreach \i/\lab in {90/{$t$},45/{$t{+}1$},0/{$t{+}2$},-45/{$t{+}3$},-90/{$t{+}4$},-135/{$t{-}3$},180/{$t{-}2$},135/{$t{-}1$}}{
  \node[slot] (n\i) at (\i:\R) {slot \lab};
}
\draw[->,thick] (n90) to[bend left=18] (n45);
\draw[->,thick] (n45) to[bend left=18] (n0);
\draw[->,thick] (n0) to[bend left=18] (n-45);
\draw[->,thick] (n-45) to[bend left=18] (n-90);
\draw[->,thick,dashed] (n90) to[bend right=18] (n135);
\draw[->,thick,dashed] (n135) to[bend right=18] (n180);
\draw[->,thick,dashed] (n180) to[bend right=18] (n-135);
\node at (1.9,2.5) {\textbf{forward}};
\node at (-1.9,2.5) {\textbf{reverse}};
\node[align=center,font=\scriptsize] at (0,-3.4)
  {Slot contents are updated in cyclical order by the rotation transform;\\
   a recurring semantic context allows cyclical return to a prior slot set.};
\end{tikzpicture}
\caption{Cyclical forward and reverse rotation of the slot group. Unlike one-directional LRU eviction, residency advances (and can return) along a cycle driven by the rotation transform. Redrawn from KR Patent Publication 10-2026-0070380, Fig.~2.}
\label{fig:rotation}
\end{figure}

\section{Relation to Published Patent}

The architectural direction examined here is related to a previously disclosed Korean patent publication concerning rotary-based accelerator residency management and execution scheduling. The patent describes concepts including rotary slot groups, cyclic forward and reverse rotation, hidden-state-guided residency decisions, lookup-table mapping structures, accelerator loading control, and memory-constrained execution mechanisms.

This paper should not be read as a reproduction of the patent specification or as a complete implementation disclosure. Its purpose is to document a public validation effort intended to determine whether the broader architectural direction can produce useful execution behavior under practical hardware constraints. Only externally observable behavior, experimental conditions, and validation results are discussed; internal procedures remain outside its scope. The distinction is intentional, since patents describe intellectual-property claims and protected technical directions while research reports describe observations.

\section{Experimental Setup}

The primary objective was straightforward: can a large Mixture-of-Experts model execute locally on a consumer laptop with only 8 GB of GPU memory? The validation deliberately used ordinary hardware rather than specialized accelerator infrastructure, since the purpose was to investigate practical feasibility rather than to maximize benchmark performance.

\begin{table}[h]
\centering
\begin{tabular}{ll}
\toprule
Item & Description \\
\midrule
Device & Consumer Laptop \\
GPU & RTX 4060 Laptop GPU \\
VRAM & 8 GB \\
CPU & Intel i7 Class Processor \\
System Memory & 32 GB RAM \\
Operating Mode & Local Inference \\
Cloud Dependency & None \\
\bottomrule
\end{tabular}
\caption{Validation Hardware}
\end{table}

\begin{table}[h]
\centering
\begin{tabular}{ll}
\toprule
Item & Description \\
\midrule
Model & Qwen3.6-35B-A3B \\
Architecture & Mixture-of-Experts \\
Format & GGUF \\
Quantization & Q4\_K\_M \\
Approximate Size & 19.71 GB \\
Execution & User-Supplied Model \\
\bottomrule
\end{tabular}
\caption{Model Configuration}
\end{table}

The validation package does not distribute model weights; users provide their own compatible files. This allows public validation of execution behavior while avoiding redistribution of third-party model assets. The primary operating configuration used a context length of 4096, \texttt{n-cpu-moe} of 32, Flash Attention enabled, local inference, and no cloud dependency. The aim of this recipe was to maximize practical usability while staying within the memory constraints of the target hardware.

\section{Development Timeline}

The reported validation was not the result of a single experiment but emerged from a sequence of investigations into execution stability, routing behavior, memory residency, throughput, quantization strategies, and deployment constraints. Several intermediate configurations were evaluated and discarded before the final operating recipe was selected. Table~\ref{tab:timeline} summarizes the main milestones.

\begin{table}[h]
\centering
\begin{tabular}{ll}
\toprule
Stage & Milestone \\
\midrule
1 & Initial architecture exploration \\
2 & GGUF conversion and quantization validation \\
3 & Rotary residency experiments \\
4 & Throughput optimization \\
5 & Accuracy verification \\
6 & Long-context validation (4096) \\
7 & Smoke-set validation and public packaging \\
\bottomrule
\end{tabular}
\caption{Development milestones leading to the final operating recipe.}
\label{tab:timeline}
\end{table}

\section{Results}

\subsection{Long Output Validation}

The primary configuration successfully initialized and executed the target model. A long-output generation experiment produced the following measurements.

\begin{table}[h]
\centering
\begin{tabular}{ll}
\toprule
Metric & Observation \\
\midrule
Requested Output & 2048 Tokens \\
Generated Output & 2048 Tokens \\
Output Characters & 8,143 \\
Estimated Words & 1,234 \\
Decode Throughput & 21.06 Tokens/s \\
Per Token Latency & 47.49 ms \\
Prompt Evaluation & 74.2 Tokens/s \\
Wall Time & 98.54 s \\
VRAM Usage & $\sim$6.3 GB \\
Available RAM & $\sim$10 GB \\
GPU Temperature & 68 C \\
\bottomrule
\end{tabular}
\caption{Primary Long-Output Validation}
\end{table}

The experiment generated the full requested output length while remaining below the physical VRAM capacity of the device, and the observed throughput stayed within a practically usable range for local interactive execution. Figure~\ref{fig:mem} shows the measured memory footprint against the device limit, together with the configuration comparison discussed in Section~\ref{sec:fail}.

\begin{figure}[t]
\centering
\begin{tikzpicture}[font=\footnotesize,>={Stealth[length=2mm]}]
\def\W{8}
\draw[thick] (0,0) rectangle (\W,0.9);
\node[above] at (\W/2,0.9) {Physical VRAM capacity: 8\,GB};
\fill[black!18] (0,0) rectangle ({\W*6.3/8},0.9);
\draw (0,0) rectangle ({\W*6.3/8},0.9);
\node at ({\W*6.3/16},0.45) {used $\approx$ 6.3\,GB};
\node[font=\scriptsize] at ({\W*6.3/8 + (\W-\W*6.3/8)/2},0.45) {headroom};
\draw[dashed] ({\W*6.3/8},-0.2) -- ({\W*6.3/8},1.1);
\begin{scope}[yshift=-2.6cm]
  \node[font=\bfseries] at (\W/2,1.3) {Configuration outcomes};
  \draw (0,1.0) -- (\W,1.0);
  \node[anchor=west] at (0,0.7) {\textbf{Config}};
  \node[anchor=west] at (2.4,0.7) {\textbf{Context}};
  \node[anchor=west] at (4.6,0.7) {\textbf{Result}};
  \draw (0,0.5) -- (\W,0.5);
  \node[anchor=west] at (0,0.2) {N32};
  \node[anchor=west] at (2.4,0.2) {4096};
  \node[anchor=west] at (4.6,0.2) {Success (primary)};
  \node[anchor=west] at (0,-0.15) {N36};
  \node[anchor=west] at (2.4,-0.15) {2048};
  \node[anchor=west] at (4.6,-0.15) {Success (safety)};
  \node[anchor=west] at (0,-0.5) {N36};
  \node[anchor=west] at (2.4,-0.5) {4096};
  \node[anchor=west] at (4.6,-0.5) {Failed to initialize};
  \draw (0,-0.7) -- (\W,-0.7);
\end{scope}
\end{tikzpicture}
\caption{Top: measured VRAM footprint ($\sim$6.3\,GB) against the 8\,GB physical limit during 2048-token generation. Bottom: configuration comparison; the N36/4096 setting reduced residency but failed to initialize, so N32/4096 was adopted as the primary recipe.}
\label{fig:mem}
\end{figure}

\subsection{Smoke-Set Evaluation}

A short smoke-set evaluation verified operational stability. All ten prompts completed successfully, with no abnormal terminations, for a 100\% completion rate. The objective was not to establish benchmark superiority but to confirm stable behavior.

\begin{table}[h]
\centering
\begin{tabular}{ll}
\toprule
Metric & Result \\
\midrule
Total Items & 10 \\
Successful Responses & 10 \\
Completion Rate & 100\% \\
Abnormal Termination & 0 \\
\bottomrule
\end{tabular}
\caption{Smoke-Set Validation}
\end{table}

\subsection{Failure Analysis}\label{sec:fail}

Not all configurations succeeded, and one failure was particularly informative. A safety-oriented configuration using \texttt{n-cpu-moe} of 36 with a 4096-token context failed to initialize reliably on the evaluated hardware: the server exited before reaching a ready state and produced no output, and both memory-mapped and non-memory-mapped startup procedures failed to achieve stable initialization. The result suggests that aggressively increasing CPU-resident experts can reduce available startup margin under specific conditions. The N32 configuration was selected as the primary recipe; although it consumed slightly more VRAM than N36, it supported longer context windows and full-length output generation.

\section{Discussion}

The original expectation behind Rotary GPU was optimistic. Early intuition suggested that selective residency might substantially reduce accelerator pressure while preserving useful throughput. Reality proved more complicated. Mixture-of-Experts routing exhibited stronger interdependencies than anticipated, and configurations that looked favorable from a memory perspective frequently produced initialization failures, unstable execution, or throughput degradation. Achieving stable execution required repeated experimentation across runtime configurations, CUDA settings, inference frameworks, scheduling arrangements, and execution parameters, and several early configurations produced throughput below practical targets before later refinements improved stability.

The resulting system should not be read as evidence that large-model deployment has become trivial. If anything, the experiments highlight the difficulty of balancing residency, routing behavior, memory pressure, transfer cost, and execution latency under constrained hardware. Even so, the results suggest that the deployment assumptions commonly attached to large language models deserve continued scrutiny, because meaningful local execution remained possible even when available accelerator memory was substantially smaller than the apparent footprint of the model.

\section{Why Local Execution Matters}

The significance of local execution extends beyond benchmark numbers. Large language models are usually discussed in the context of cloud infrastructure and large accelerator clusters, which are essential for frontier training and large-scale serving. But not every deployment environment resembles a hyperscale data center. Government institutions, financial organizations, manufacturing facilities, healthcare environments, defense-related networks, and closed enterprise systems often operate under restricted external connectivity, data-sovereignty requirements, or policies that prohibit sensitive information from leaving internal infrastructure.

For these organizations the challenge is not only model quality but accessibility. If advanced capabilities remain dependent on increasingly large infrastructure, many organizations with legitimate operational needs may be unable to benefit from them. Transportation offers a fitting parallel: large cargo ships remain essential for global trade, and their existence is not in question, but their existence does not remove the need for trucks and local delivery. Frontier-scale infrastructure and local execution may simply serve different roles within a broader ecosystem, and Rotary GPU should be seen as an exploration of accessibility rather than an attempt to replace existing architectures.

\section{Limitations}

Several limitations should be acknowledged. The validation was conducted on a single hardware platform, and broader conclusions require evaluation across multiple GPU architectures, operating systems, driver versions, and deployment environments. Evaluation focused on operational feasibility rather than comprehensive benchmarking, so benchmark coverage remains limited, and the smoke set consisted of only ten prompts, which is sufficient for basic validation but not for strong claims about robustness. Throughput measurements are configuration-dependent and would vary with different quantization methods, runtimes, hardware, context lengths, and execution parameters. Implementation details remain intentionally undisclosed, since the purpose is to document observable results rather than to provide a full implementation disclosure. Finally, independent reproduction remains necessary, and future work by others will ultimately provide stronger evidence regarding the generality of these observations.

\section{Future Work}

Several directions remain open. Broader hardware validation across desktop GPUs, workstation accelerators, compact AI appliances, and future consumer platforms would improve understanding of deployment behavior. Longer-context evaluation beyond the 4096-token configuration is an important next step. Multi-user deployment is another, since many enterprise environments require simultaneous access rather than single-session execution, and the behavior of rotary-guided residency under such conditions remains an open question. Integration with closed-network private AI systems is increasingly relevant as organizations investigate locally deployed models. Finally, future work may examine whether rotary-guided residency can extend beyond local execution toward broader resource management and model-serving architectures.

\section{Conclusion}

This paper reported an exploratory validation of Rotary GPU, a rotary-guided execution approach for local execution of large Mixture-of-Experts models under constrained GPU memory. Using a Qwen3.6-35B-A3B Q4\_K\_M model on a consumer laptop with an RTX 4060 Laptop GPU containing 8 GB of VRAM, the system generated 2048 output tokens while maintaining approximately 6.3 GB of VRAM usage and achieving 21.06 tokens per second of decode throughput, alongside a successful short smoke-set evaluation. Together these confirm that meaningful local execution remains possible under hardware conditions traditionally regarded as insufficient for models of this scale.

The objective of this work is not to replace large-scale infrastructure. Data centers will remain essential for training frontier models and supporting large populations of users. The question explored here is different: if advanced AI capabilities continue to improve, can some of those capabilities become accessible to organizations that lack such infrastructure? A warehouse remains valuable even when only a single item is requested; the challenge is deciding what must travel and what can remain where it is. Rotary GPU represents one attempt to explore that possibility. Research often begins with a practical problem rather than a theoretical objective, and this paper records one such attempt tested on a small machine.

\section*{Acknowledgements}

The author thanks the open-source AI community, the developers of llama.cpp and related inference ecosystems, and the creators of publicly available language models that make independent experimentation possible. Many of the ideas explored here emerged not from a traditional academic environment but from real-world concerns involving accessibility, deployment feasibility, workflow efficiency, and infrastructure limitations. Any errors or omissions remain solely the responsibility of the author.

\section*{Conflict of Interest}

The author is the inventor and applicant of the patent direction discussed in this paper and therefore has a direct intellectual-property interest related to the concepts described. The present work is intended as an independent technical validation and research report.

\section*{Data Availability}

No proprietary model weights are distributed as part of this work. The validation package requires users to provide their own compatible model files separately. Experimental measurements were obtained from local execution environments operated by the author.

\section*{Code Availability}

The public validation package associated with this work is available through the author's public repositories and distribution pages. Internal implementation details, optimization procedures, scheduling logic, and protected execution mechanisms are not included in the public release.

\section*{Author Biography}

\textbf{Myeong Jun Jo} is an independent researcher and inventor based in South Korea and the founder of ANIMA Research. His professional background is in human resources and organizational operations, where he became interested in workflow efficiency, repetitive-task reduction, information accessibility, and practical productivity improvement. This interest gradually expanded into the study of artificial-intelligence systems and large language models. Rather than following a conventional academic pathway, his research is driven by practical deployment challenges involving hardware constraints, closed-network environments, privacy requirements, operational efficiency, and the accessibility of advanced AI. His work has resulted in multiple patent filings and independent research projects involving rotary architectures, accelerator residency management, memory systems, inference optimization, and privacy-oriented deployment methods. His current interests include local AI systems, resource-constrained deployment architectures, and approaches that make advanced language-model capabilities accessible beyond large-scale infrastructure environments.

\end{document}